\documentclass[prd,twocolumn,aps,showpacs,nofootinbib,nobibnotes,superscriptaddress]{revtex4}
\usepackage{amssymb}
\usepackage{graphicx}
\usepackage{bm}

\begin{document}

\title{Detectability of Gravitational Waves from Phase Transitions}

\date{\today}

%\date{\today~~KSUPT-08/1}
\author{Tina Kahniashvili}
\email{tinatin@phys.ksu.edu} \affiliation{Department of Physics,
Kansas State University, 116 Cardwell Hall, Manhattan, KS 66506,
USA}\affiliation{CCPP, New York University, 4 Washington Plaza,
New York,  NY 10003, USA} \affiliation{Department of Physics,
Laurentian University, Ramsey Lake Road, Sudbury, ON P3E 2C6,
Canada} \affiliation{National Abastumani Astrophysical
Observatory, 2A Kazbegi Ave, Tbilisi, GE-0160, Georgia}
\author{Arthur Kosowsky}
\email{kosowsky@pitt.edu}
\affiliation{Department of Physics and
Astronomy, University of Pittsburgh, 3941 O'Hara Street,
Pittsburgh, PA 15260 USA}
\author{Grigol Gogoberidze}
\email{gogober@geo.net.ge} \affiliation{Department of Physics,
Kansas State University, 116 Cardwell Hall, Manhattan, KS 66506,
USA}\affiliation{Centre for Plasma Astrophysics, K.U.\ Leuven,
Celestijnenlaan 200B, 3001 Leuven, Belgium}\affiliation{National
Abastumani Astrophysical Observatory, 2A Kazbegi Ave, Tbilisi,
GE-0160, Georgia}
\author{Yurii Maravin}
\email{maravin@phys.ksu.edu} \affiliation{Department of Physics,
Kansas State University, 116 Cardwell Hall, Manhattan, KS 66506,
USA}

\begin{abstract}
Gravitational waves potentially
represent our only direct probe of the universe when it was less than
one second old. In particular,
first-order phase transitions in the early universe can generate a stochastic
background of gravitational waves which may be detectable today.
We briefly summarize the physical sources of gravitational radiation from phase transitions
and present semi-analytic expressions for the resulting
gravitational wave spectra from three distinct realistic sources:
bubble collisions, turbulent plasma motions, and inverse-cascade
helical magnetohydrodynamic turbulence. Using phenomenological parameters
to describe phase transition properties, we determine the region of parameter space
for which gravitational waves can be detected by the
proposed Laser Interferometer Space Antenna. The electroweak phase
transition is detectable for a wide range of parameters.
\end{abstract}

\pacs{98.70.Vc, 98.80.-k}

\maketitle

\section{Introduction}

Since gravitational waves propagate freely through the universe
after being generated, their detection provides a powerful test
of the very early universe. Various mechanisms that generate such
gravitational waves have been discussed: quantum fluctuations
during \cite{inflation} and shortly after inflation
\cite{end-inflation}; bubble wall motion and collisions during
phase transitions \cite{bubble,kos1,ktw92a,kt93,kos2}; cosmic
strings and defects, including primordial black holes
\cite{strings}; cosmological magnetic fields
\cite{magnet,cdk04,cd06},
 and plasma turbulence \cite{kmk02,dolgov,kgr05,gkk07}.
This paper focuses on gravitational
waves generated during phase transitions and study the possibility
of direct detection by the planned Laser Interferometer Space Antenna (LISA),

Near-future data from the Large Hadron Collider
will for the first time have the ability to probe in detail
physics at the electroweak energy scale. This physics determines
the nature of the electroweak phase transition in the early
universe, when the primordial plasma went from an electroweak-symmetric
state to a broken state with distinct electromagnetic and weak interactions.
The phase transition took place when the primordial plasma had a
temperature on the order of 1 TeV.
Intriguingly, the Hubble frequency $H_*$ at this epoch, redshifted
to today, falls in the lower end of the detection range for
LISA, which ranges from $10^{-4}$ to $10^{-1}$ Hz \cite{lisa}.
If gravitational waves are generated during the
electroweak phase transition, their characteristic frequency will be
related to this Hubble frequency: remarkably, gravitational waves may provide us
with an alternate route to probing electroweak physics.

Having gravitational waves at detectable frequencies is not
sufficient for detecting the electroweak phase transitions: the
source of gravitational waves must be sufficiently strong to
produce radiation with a detectably large amplitude today
\cite{detection,nicolis,gs06,grojean}. If the phase
transition is first order, the latent heat of the phase transition partly
is transferred into kinetic energy of the walls of
expanding bubbles of the broken phase. If these expanding
bubbles contain a large enough amount of energy, they
will produce a gravitational wave background with
a detectably large amplitude today \cite{bubble}. Models of
the electroweak phase transition based on standard-model particle
physics do not produce observable gravitational wave signals since
they are not first order for allowed values of the Higgs mass \cite{stdmodelweak}, but
common extensions of the standard model, including supersymmetry
and extra dimensions,
can produce much stronger phase transitions \cite{strong}.

Bubbles of broken phase in a first-order phase transition expand
and percolate to convert the entire universe to the low-energy phase.
The kinetic energy in the bubble walls eventually thermalizes, but
prior to that the bubbles act to stir the primordial plasma, plausibly
generating Kolmogoroff turbulence cascading from the turbulence
scale to a much smaller scale where the kinetic energy turns into thermal
energy via viscous heating. The turbulent eddy motions
can also be a potent source of gravitational waves if the energy
input is large enough \cite{kmk02}. If a small magnetic seed field with
nonzero helicity is generated at the phase transition \cite{helicity,ste08},
magnetohydrodynamic effects can generate an
inverse cascade, transferring energy to scales substantially larger
than the stirring scale \cite{gkk07} and resulting in a detectably
large signal \cite{kgr08}, and additionally can give
the radiation a non-zero circular polarization \cite{kgr05} which
is also directly detectable \cite{seto}. So gravitational wave generation
from an early universe phase transition can be decomposed into three distinct
sources: expanding bubbles of the broken phase, hydrodynamic turbulence
stirred by the colliding bubbles, and an inverse cascade due to the
amplification of seed magnetic fields.

Previous papers have computed the gravitational wave spectra for each of these
sources individually. Here we consider their combined spectrum. The
detailed shape and amplitude of the spectrum depends on fundamental
properties of the phase transition: its energy scale, bubble nucleation rate,
latent heat, efficiency of converting latent heat into plasma motions,
and mean helicity of seed magnetic fields. We summarize how the total
power spectrum depends on these parameters, then evaluate the regions of
parameter space for which the relic gravitational radiation from the
phase transition is detectable with LISA, displaying the results as contour
plots in the parameter space. While the bubble collisions produce detectable
gravitational radiation for a very limited range of parameters, adding the
radiation from turbulence and an MHD inverse cascade greatly widens the
range of detectable phase transitions.

In the following Section,  we give a brief overview of the
physics of phase transitions and the corresponding phenomenological parameters.
In Sec.~III, we express the gravitational wave spectra in terms of
these phenomenological parameters, and show how the total gravity wave
spectrum varies with the parameters. Sec.~IV compares these spectra with the
projected LISA sensitivity curve for stochastic backgrounds, displaying
detectability regions in the space of phase transition parameters.

\section{Modeling Gravitational Radiation}

\subsection{General Remarks on
 Gravitational Waves Generation}

 Gravitational waves  generated during phases transitions
 by stochastic sources have an energy density just after the phase transition of \cite{m00}
\begin{eqnarray}
&&\rho_{\rm GW} ({\bf x}) = \frac{1}{32\pi G} \langle
\partial_t h_{ij}({\mathbf x},t) \partial_t h_{ij} ({\mathbf
x},t)\rangle   = ~~~~~~\nonumber \\
&&~= \frac{G}{2\pi } \int {\rm d}^3 {\bf x}^\prime {\rm d}^3 {\bf
x}^{\prime \prime} \frac{\langle
\partial_t S_{ij}({\mathbf x}^\prime,t^\prime) \partial_t
S_{ij}({\mathbf x}^{\prime \prime},t^{\prime \prime}) \rangle
}{|{\bf x}- {
\bf x}^\prime| |{\bf x}- {\bf x}^{\prime \prime}|}.
\label{eq:04}
\end{eqnarray}
where time variables with primes represent the light-cone
combination $t^{\prime}=t-|{\bf x}- {\bf x}^{\prime}|$, $i$ and
$j$ are spatial indices (repeated indices are summed), the source
$S_{ij}({\mathbf x}, t)= T_{ij} ({\mathbf x}, t) - \delta_{ij}
T^k_k ({\mathbf x}, t) /3$ is the traceless part of the
stress-energy tensor $T_{ij}$, $G$ is the gravitational
constant,  and we use natural units with $\hbar = c = 1$. This
expression assumes that the duration of the phase transition is
short enough to neglect the cosmological expansion during
gravitational wave production. Since the phase transition lasts
for a finite duration, we can consider the total radiation field
to be an incoherent sum of radiation from many sources, each with
a size corresponding to the light travel distance during the
phase transition. With this approximation, we can apply the
far-field approximation for each individual source region, $x \gg
d$ where $d$ is the size of the source region; in this region the
gravitational waves are the only metric perturbations \cite{W},
allowing replacement of $|{\bf x}-{\bf x}^\prime|$ by $|\bf x|$
in Eq.~(\ref{eq:04}).

Generated gravitational waves propagate freely through the
expanding universe until today. Their wavelengths simply scale
with the scale factor $a$ of the universe, while their total
energy density evolves like $a^{-4}$ and their amplitude decays
like $a^{-1}$. Following Ref. [27], we use the gravitational wave
spectral energy density $\rho_{GW}(\omega)$ per logarithmic
angular frequency $\omega$, defined as the root-mean-square
average of $\rho_{GW}({\bf x},\omega)$ over spatial positions,
and form the spectral energy density parameter
$\Omega_{GW}(\omega) = \rho_{GW}(\omega)/\rho_c$ with the
critical energy density $\rho_c = 3H_0^2/8\pi G$. Changing to
linear frequency $f=\omega/2\pi$, a characteristic strain
amplitude is conventionally defined as
\begin{equation}
h_c(f) = 1.3 \times 10^{-18} \left(\frac{1\,{\rm Hz}}{f}\right)\left[h_0^2\Omega_{\rm GW}(f)\right]^{1/2}
\label{hcdef}
\end{equation}
where $h_0$ is the current Hubble parameter $H_0$ in units of 100
${\rm km}\,{\rm sec}^{-1}{\rm Mpc}^{-1}$. We will present our
results in terms of the gravitational wave amplitude $h_c$ vs.
linear frequency $f$ measured today.

The temperature at which the phase transition occurs determines
the corresponding Hubble frequency, which is given today by
\cite{kmk02}
\begin{equation}
f_{\rm H} = 1.6 \times 10^{-5} \,{\rm Hz } \left(
\frac{T_*}{100\,{\rm GeV}} \right) \left( \frac{g_*}{100}
\right)^{1/6} \label{hubble-frequency}
\end{equation}
where $g_*$ is the number of relativistic degrees of freedom
at the temperature $T_*$; for standard model degrees of
freedom, $g_*\sim 106.75$ as $T\rightarrow\infty$. To obtain this expression,
begin with the Hubble parameter at temperature $T_*$, given by
\begin{equation}
H^2_* = \frac{8\pi^3 T_*^4}{90M_{\rm Pl}^2} \label{H-star}
\end{equation}
then rescale $H_*$
to its present value with the ratio between the scale factors
$a_\ast$ corresponding to the Universe temperature $T_\ast$
and $a_0$ today:
\begin{equation}
\frac{a_\ast}{a_0} = 8 \times 10^{-16} \left(\frac{100\,{\rm GeV}}{T_\ast}\right)
\left(\frac{100}{g_\ast}\right)^{\frac{1}{3}}. \label{a-ratio}
\end{equation}
Note that an extra factor of approximately 3.6 occurs for $f_{\rm H}$
in Ref.~\cite{gs07} (Eq.~10) due to approximating
 $H_*^2 \simeq G\rho_{\rm tot}$.

It can be shown from general
physical principles that since the gravitational wave generation
process is causal,  the peak frequency must be greater than or equal
to the Hubble frequency, $f_{\rm peak} \geq f_{\rm
H}$ \cite{bubble}. Independent of the nature of the source, the
gravitational wave spectrum is the same at low frequencies,
$\Omega_{\rm GW}(f)\propto
(f/f_{\rm peak})^{3}$ (for $f\ll f_{\rm peak}$)
\cite{ktw92a,kmk02,nicolis,cd06,gkk07,cds07}. The gravitational
spectrum behavior at higher frequencies is determined by the
specific source features.

\subsection{Phase Transition Model Parameters}

A first-order phase transition is generically described by
several parameters: (i) $\alpha \equiv \rho_{\rm vac} / \rho_{\rm
thermal}$, the ratio of the vacuum energy
 associated with the phase transition to the thermal density of
the Universe at the time (which characterizes the strength of the
phase transition); (ii) $\kappa$, an efficiency factor which
gives the fraction of the available vacuum energy which goes into
the kinetic energy of the expanding bubble walls, as opposed to
thermal energy; (iii) $\beta^{-1}$, which sets the
characteristic time scale for the phase transition; (iv) $v_b$,
the velocity of the expanding bubble walls, which set the
characteristic length scale of the phase transition; (v) $T_*$,
the temperature at which the phase transition occurs.
In any first-order phase transition, the
characteristic bubble nucleation rate per
unit volume is generically \cite{tww92}
\begin{equation}
\Gamma = \Gamma_0 e^{\beta t} \label{gamma_def},
\end{equation}
which uses the constant and linear terms of a Taylor expansion
of the bubble nucleation action.
General considerations confirmed by numerical calculations
show that the largest bubbles reach a size
of order $\beta^{-1} v_b$ by the end of the phase transition
\cite{kt93}, where $v_b$ is the bubble expansion velocity,
assuming the bubbles remain spherical as they expand. In general,
$\beta$ is expected to be of the order $4\ln(m_{\rm Pl}/T) H
\simeq 100 H$ for a Hubble rate $H$ \cite{tww92}.

The fundamental symmetry breaking mechanism which drives the phase
transition  determines some effective potential for bubble
nucleation \cite{col77}. The difference in energy density between
the two phases and the bubble nucleation rate are both determined
by this mechanism. Thus the parameters $T_*$, $\beta$, and
$\alpha$ are all determined directly by the underlying physics.
On the other hand, the bubble velocity  $v_b$ and the fraction of
kinetic energy into the bubbles $\kappa$ depend on the detailed
microphysics involved in the bubble propagation through the
relativistic plasma and are not determined from general
properties of the effective potential. In general, the larger the
vacuum energy density driving phase transition, the higher the
bubble wall velocity $v_b$ \cite{tww92,kos1}.

\section{Sources of Gravitational Radiation}

\subsection{Bubble collisions}

Bubbles walls can propagate outwards through a relativistic plasma
via two modes, detonation and deflagration
\cite{ste82}. In the case of detonation, the bubble walls are thin compared
to the bubble radius with velocity
is \cite{ste82}
\begin{equation}
v_b(\alpha) = {1/\sqrt{3} + (\alpha^2 + 2\alpha/3)^{1/2} \over
1+\alpha} \label{vb_alpha}
\end{equation}
and an approximate form of $\kappa$ is given \cite{kos1}
\begin{equation}
\kappa(\alpha) = {1\over 1+A\alpha} \left[A\alpha + {4\over
27}\left(3\alpha\over 2\right)^{1/2}\right] \label{kappa_alpha}
\end{equation}
with $A=0.715$. If the bubbles propagate as a deflagration front,
the walls are thick and have lower energy density. However, it has been argued
\cite{kf92} that for relativistic plasma, instabilities in the
bubble shape will accelerate the bubble walls and the hydrodynamic deflagration
expansion mode is unstable to becoming a detonation. For this
reason, Ref. \cite{kos1} assumed that Eqs.~(\ref{vb_alpha}) and
(\ref{kappa_alpha}) hold, and we will also make that assumption here.
(For  unusual cases with very strong detonation and $\alpha\gtrsim 1$,
the following formulas must be corrected.)
Both modes of
bubble propagation recently have been re-addressed in Ref.~\cite{cds07},
with results slightly different from those of Ref.~\cite{kos1}.

The peak frequency of the gravitational wave spectrum is of fundamental
importance for issues of detectability. The bubble nucleation
 time scale is always determined by $\beta^{-1}$, while the
 largest bubble size depends also on the bubble wall velocity, $l_0 = v_b \beta^{-1}$.
 What scale determines the peak frequency of the gravitational waves?
 When the bubble wall velocity $v_b$ approaches the speed of light,
 $l_0^{-1} \simeq \beta$, and both characteristic frequency scales are the same.
 In Refs.~\cite{bubble,ktw92a,kos1}, where the bubble wall velocity
is assumed to be the speed of light, the peak frequency is given
by \cite{kos1,nicolis,cds07}
\begin{equation}
f_{\rm peak}^{\rm BC} = 0.3  \left(\frac{\beta}{H_\star} \right)
 f_{\rm H}. \label{f-peak-bubble}
\end{equation}
Note the angular peak frequency at the moment of emission
$\omega^{\rm BC}_{{\rm peak}\, \star}$ is set by $0.3 k_0 v_b
\simeq 2 \beta$ (where $k_0 = 2\pi/l_0$).  To understand the peak
frequency of the radiation from bubble walls with lower velocity,
it is useful to exploit the direct analogy with sound wave
generation. The typical source velocity compared to the velocity
of the induced waves determines the characteristic peak frequency
\cite{G}. If the characteristic source velocity, given by the
ratio between the source's characteristic length and time scales,
is less than the group velocity of the induced waves, then the
frequency is determined by the source time scale; in the opposite
case, the characteristic length scale determines the peak
frequency. Applied to gravitational waves with group velocity
$v=1$, the source time scale always determines the peak radiation
frequency, since by causality its characteristic velocity can
never exceed the speed of light.
%will have its peak radiation frequency determined by
%the source time-scale, for the source with $v_{\rm source}
%\rightarrow 1$ the peak frequency can be also determined through
%the source characteristic spatial scale \cite{cds07},

 For the realistic values $\beta/H_*=100$
and $g_* =100$, the peak frequency from bubble collisions $f^{\rm
BC}_{\rm peak} \simeq 5 \times 10^{-4}~{\rm Hz}$ \cite{nicolis} is
close to the peak of LISA sensitivity at $2 {\rm mHz}$ if $T_*
\simeq 100{\rm GeV}$. The corresponding peak amplitude of the
gravitational radiation today is \cite{kos1}
\begin{eqnarray}
h_c(f^{\rm BC}_{\rm peak}) &=& 1.8 \times 10^{-16} \kappa
\left(\frac{\alpha}{\alpha +1}\right)
\left(\frac{H_\star}{\beta}\right)^2
\left(\frac{100}{g_\star}\right)^{\frac{1}{3}} \nonumber \\
&\times &\left(\frac{100{\rm
Gev}}{T_\star}\right)\left(\frac{v_b^5}{v_b^3+0.24}\right)^{\frac{1}{2}}.
\label{h-peak-bubble}
\end{eqnarray}
The spectrum shape at high frequencies ($f \gg f_{\rm H}$) is
 $h_c(f) \propto f^{-2}$ \cite{cds07}, very close to the scaling
 $h_c(f)
\propto f^{-1.9}$ found numerically in Ref.~\cite{kos1}.

\subsection{Turbulence}

Once the bubbles expand and percolate, a significant amount of
their kinetic energy will be converted to turbulent bulk motions
of the primordial plasma \cite{ktw92a}. As the phase transition
ends, far more small bubbles have been nucleated than large ones,
but the energy density in the large ones dominates the total
energy density \cite{tww92}. Therefore, a reasonable approximation is that
the turbulence energy is injected on a stirring scale $l_0$
corresponding to the size of the largest bubbles, and that
that  the stirring will last for roughly $\tau_{\rm
stir}=\beta^{-1}$, comparable to the duration of the
phase transition.  We make the conservative assumption that
the duration of the turbulence is also $\tau_{\rm stir}$. The turbulence
can actually last substantially longer than this; see Ref.~\cite{kmk02} for
a detailed discussion. After  generation, the turbulence kinetic
energy  cascades from larger to smaller scales, stopping at a
viscous damping scale
$l_D$ where the turbulence energy is dissipated into heat. We
define the parameter $\gamma \equiv l_0 H_\star \simeq v_b (H_\ast /
\beta)$ which determines the number ($N_{\rm eddy} \simeq
\gamma^{-3}$) of turbulent eddies within the Hubble radius. In terms
of the parameter $\gamma$, Eq.~(\ref{f-peak-bubble}) reads
$f^{\rm BC}_{\rm peak} = 0.34 f_{\rm H} v_b /\gamma$.

We associate a characteristic velocity perturbation $v_0$ with the fluid
velocity of the
largest eddies at the stirring scale $l_0$. This velocity determines
the turbulent motions' Mach number $M\simeq v_0$. It is easy to estimate
$v_0$ through relating the kinetic energy density of plasma ${\rm
w} v_0^2/(1-v_0^2)$ (where ${\rm w}=p+\rho$ is enthalpy of
plasma) and the vacuum energy $\rho_{\rm vac}$ using ${\rm w} =
4\alpha\rho_{\rm vac}/3$. This leads to \cite{nicolis}
\begin{equation}
v_0 = \sqrt{\frac{3\kappa\alpha}{4 + 3\kappa \alpha}}.
\label{v-turbulence}
\end{equation}
Accounting for Eq.~(\ref{kappa_alpha}), $v_0$ can be expressed
uniquely in terms of $\alpha$ and in the case of realistic phase
transitions, when  $\alpha < 1$, $v_0 \simeq \alpha^{3/4}
(3/2)^{1/4}/3$, and of course is less than the bubble wall
expansion speed $v_b$.  For $\kappa \alpha \simeq 1$
corresponding to a strongly first order phase transition, $v_0 \simeq
0.65$.

Figure~1 of Ref. \cite{gkk07} shows that it is safe to adopt the
aero-acoustic limit $k\rightarrow 0$ in the gravitational wave generation formalism
for turbulent sources, provided only that the Mach number
$M \lesssim 1$. Employing this simplifying assumption and
adopting the results of Ref. \cite{gkk07},
the peak frequency of gravitational waves is given by
\begin{equation}
f^{\rm T}_{\rm peak}= \left( \frac{M}{v_b}\right) f_{\rm H}
\left(\frac{\beta}{H_\star}\right) \simeq 0.3
\left(\frac{M}{v_b}\right) f^{\rm BC}_{\rm peak}.
\label{peak-hydro-turbulence}
\end{equation}
For typical parameters $\beta=100H_*$, $T_* = 100\,{\rm GeV}$, $g_*=100$,
and $\alpha =0.5 $ (which correspond to $\kappa=0.36$,
$v_b=0.89$ and  $v_0=0.34$), the peak frequency is $0.4$ mHz, within
the LISA range.

At frequencies high compared to the characteristic damping
frequency, $f \gg f_D \equiv f^{\rm T}_{\rm peak} R^{1/2}$ with
$R$ is the turbulence Reynolds number, the gravitational wave
strain spectrum $h_c(f)\sim (f/f_D)^{1/2}\exp(-(f/f_D)^2)$
possesses a sharp exponential cutoff. Near the peak frequency, for
$f^{\rm T}_{\rm peak} < f < f_D$, the spectrum scales like
$h_c(f)\simeq(f/f^{\rm T}_{\rm peak})^{-13/4}$ \cite{gkk07}. The
peak frequency of the radiation spectrum from bubble collisions
for the parameters considered here is lower than the peak
frequency due to turbulence, $f^{\rm BC}_{\rm peak} < f^{\rm
T}_{\rm peak}$, and the bubble collision spectrum falls only as
$(f/f^{\rm BC}_{\rm peak})^{-2}$ at frequencies above the peak,
but even at frequencies substantially greater than $f^T_{\rm
peak}$, the total gravitational wave spectrum is not dominated by
the bubble collision source, due to the lower amplitude of
gravitational waves induced by collisions.

The gravitational wave amplitude from turbulence can be
conveniently estimated analytically using the aeroacoustic
approximation (see Refs. \cite{gkk07,kgr08} for details of
numerical computations), giving
\begin{eqnarray}
h_c(f^{\rm T}_{\rm peak}) &=&3.3 \times 10^{-16} M^{\frac{3}{2}}
v_b^2 \left(\frac{H_\star}{\beta}\right)^{2}
\left(\frac{100}{g_\star}\right)^{\frac{1}{3}} \nonumber\\
&\times&  \left(\frac{100\,{\rm GeV}}{T_\star}\right) \simeq  1.9
M^{\frac{3}{2}} \left(\frac{\alpha +1}{\kappa\alpha}\right)
\nonumber\\
&\times& \left(\frac{v_b}{v_b^3+0.24}\right)^{-\frac{1}{2}} h_c(f^{\rm
BC}_{\rm peak}) \label{h-c-peak-turbulence}
\end{eqnarray}
and for $\alpha \simeq 0.5 $ is approximately 3 times larger
that the peak amplitude coming from bubble collisions.

\subsection{MHD Inverse Cascade Turbulence}

Several models lead to generation of a cosmological helical
magnetic field during phase transitions \cite{helicity,ste08}. In
such a case, turbulence develops in a magnetized plasma, which is
qualitatively different than an unmagnetized plamsa. In
particular, an inverse cascade which generates magnetized
perturbations on larger scales than the stirring scale can occur.
To model such turbulence, we assume that the hydrodynamic and
magnetic Reynolds numbers are much greater than unity on scales
$\sim l_0$ (which is simply the condition for turbulence to
develop); we also assmue that the dynamics of magnetohydrodynamic
turbulence is dominated by Alfv\'en waves for which the magnetic
and kinetic energy densities are in approximate equipartition
\cite{B03}. We also assume a small initial magnetic helicity,
$\zeta_\ast \equiv {\mathcal H}_M(t_\ast)/ [2 \xi_M(t_\ast)
{\mathcal E}_M(t_\ast)] \ll 1$, of a magnetic field with the
magnetic-eddy correlation length $\xi_M(t) \equiv \int
E^M(k,t)k^{-1} dk/{\mathcal E}_M(t)$ and with total magnetic
energy ${\mathcal E}_M(t)=\int E^M\!(k,t) dk$ and helicity
${\mathcal H}_M(t)=\int H^M\!(k,t) dk$. From the stirring scale,
the direct cascade proceeds via a Kolmogoroff-like model
\cite{my75,K64} and the following inverse cascade stage adopts
the MHD turbulence model of Refs.\ \cite{BM99,CHB05}. Then it can
be shown that the induced gravitational wave spectrum always
peaks at the Hubble frequency $f^{\rm MHD}_{\rm peak}= f_{\rm H}$
\cite{kgr08}, and compared to the turbulence peak frequency it is
shifted to lower frequencies by a factor $(v_b/M)(H_\ast/\beta)$.

The amplitude is larger compared to hydrodynamic turbulence case as
\cite{kgr08}
\begin{eqnarray}
h_c(f^{\rm MHD}_{\rm peak}) &\simeq &32 \times \zeta_\ast^{9/8}
\left(\frac{\gamma}{0.01}\right)^{-3/4} M^{3/4} h_c(f^{\rm
T}_{\rm peak}) \nonumber \\
&= &\left(\frac{M}{v_b}\right)^{3/4} \zeta_\ast^{9/8}h_c(f^{\rm
T}_{\rm peak}). \label{h-peak-MHD}
\end{eqnarray}
This increase in amplitude is because the duration time of the
inverse cascade is longer than the duration of the direct cascade.
No efficient dissipation mechanisms exist at the largest scales,
so the cascade will stop at a scale $\xi_M = 1/k_S$ either when
the cascade time scale $\tau_{\rm cas}$ reaches the expansion
time scale $H_*^{-1}$ or when the characteristic length scale
$\xi_M(t)$ reaches the Hubble length $H_*^{-1}$. These conditions
are $\zeta_*^{-1/2}l_S^2/v_0 l_0 \leq H^{-1}_*$ or
$l_S=2\pi/k_S\leq H^{-1} _*$ (the cascade time is scale dependent
and maximal  at $k=k_S$). Defining $\gamma \leq 1$, it is easy to
see that the first condition is fulfilled first, and consequently
$k_0/k_S \leq (v_0 /\gamma)^{1/2} \zeta_\ast^{1/4}$. To have an
inverse cascade requires $k_0/k_S \geq 1$, leading to a
constraint on initial helicity $\gamma \leq M \zeta_\ast^{1/2}$.
In addition, these gravitational waves will be circularly
polarized since they are induced by parity-violating stochastic
sources \cite{kgr05}. If alternately we use the inverse cascade
model of Ref.~\cite{jedamzik}, the peak frequency of the
gravitational wave spectrum remains unchanged while the peak
amplitude is doubled \cite{kgr08,klgr08a}.
\begin{figure}
\includegraphics[width=6.4cm]{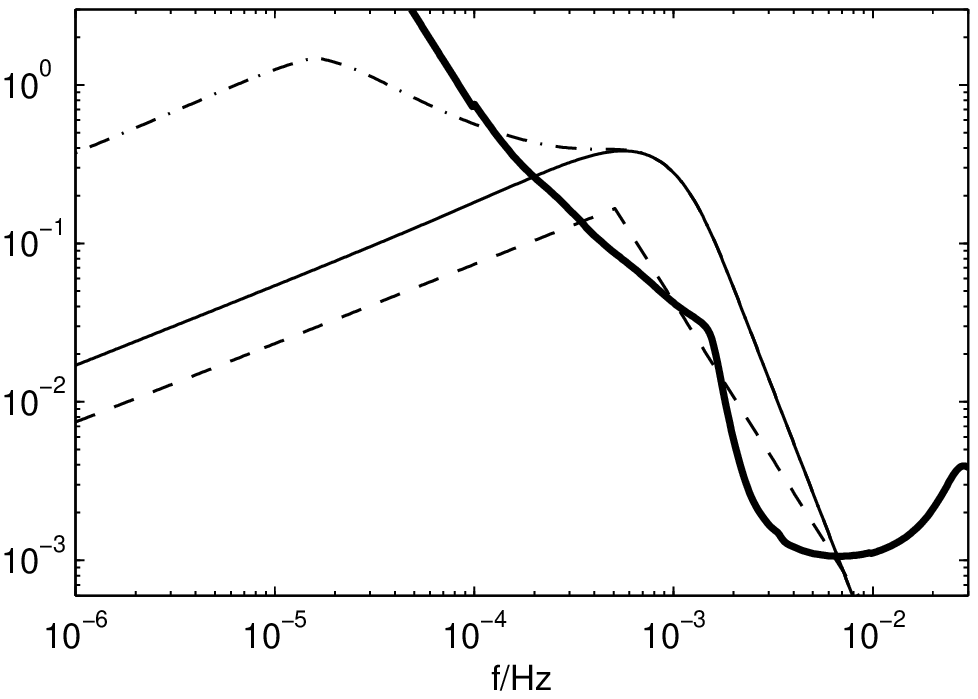}
\caption{The spectrum of gravitational radiation for a first-order
phase transition with $g_*=100$, $T_*=100\,{\rm GeV}$,
$\alpha=0.5$ and $\beta=100H_\star$, from bubble collisions (dash
line), hydrodynamic turbulence with zero helicity (solid line)
and MHD turbulence with $\zeta_*=0.15$ (dash-dotted line). The
bold solid line corresponds to the 1-year, 5$\sigma$ LISA
sensitivity curve \cite{curve}, including the confusion noise
from white dwarf binaries \cite{whitedwarfs}.} \label{fig1}
\end{figure}

\begin{figure}
\includegraphics[width=6.4cm]{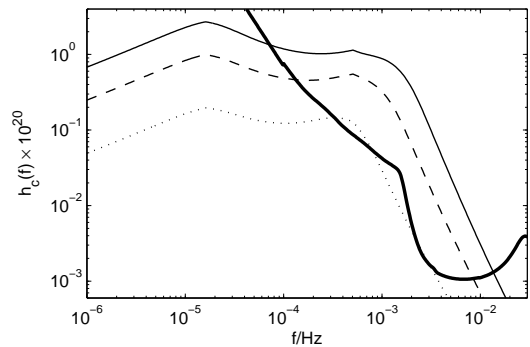}
\caption{The total spectrum of gravitational radiation (including
there sources of gravitational radiation bubble collisions,
hydro-turbulence, and MHD turbulence)  for $g_\ast=100$,
$T_\ast=100$ GeV, $\zeta_\ast=0.1$, $\beta=100 H_\ast$, and three
different values of $\alpha$: $\alpha=1$ (solid line),
$\alpha=0.5$ (dash line) and $\alpha=0.2$ (dotted line),  with the LISA sensitivity curve..}
\label{fig:spectrum-alpha}
\end{figure}

\begin{figure}
\includegraphics[width=6.4cm]{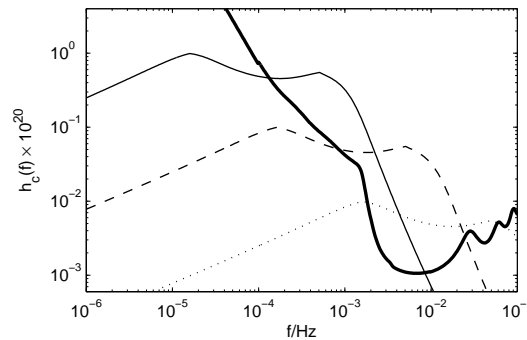}
\caption{The total spectrum of gravitational radiation  for
$g_\ast=100$, $\alpha=0.5$, $\beta=100 H_\ast$, $\zeta_\ast =0.1$,
and three different temperature values, $T_\ast=100\, {\rm GeV}$
(solid line), $T_\ast=1\, {\rm TeV} $ (dash line), and
$T_\ast=10\, {\rm TeV}$ (dotted line), with the LISA sensitivity
curve.} \label{fig:spectrum-T}
\end{figure}

\subsection{Total Gravitational Wave Spectrum}

The total gravitational wave background from a first-order phase transition
is a sum of three terms corresponding to the three distinct sources
discussed here. Our results are expressed in terms of the effective strain
spectrum $h_c(f)$ in terms of the phase transition temperature
$T_\ast$, the number of effective relativistic degrees of freedom $g_\ast$,
the ratio of vacuum energy to thermal energy $\alpha$, the bubble nucleation
time scale $\beta$, and the initial magnetic helicity
parameter $\zeta_\ast$. All others relevant parameters appearing
in the analytical approximations, such as $\kappa$, $v_b$, $l_0$,
and $M$ can be computed in terms of these fundamental parameters
using simple and well-motivated assumptions. We fix $g_* = 100$, and we do not study the
dependence on this parameter since it is very weak and the actual value
of $g_*$ will not vary more than a factor of a few from this nominal value
($g_*=106.75$ for standard model degrees of freedom at temperatures large
compared to the W and top quark masses).   We choose
$\zeta_*$ small enough to insure that the resulting magnetic turbulence
can be modeled as an inverse cascade modeling (see Ref.~\cite{kgr08} for details).

\begin{figure}
\includegraphics[width=6.4cm]{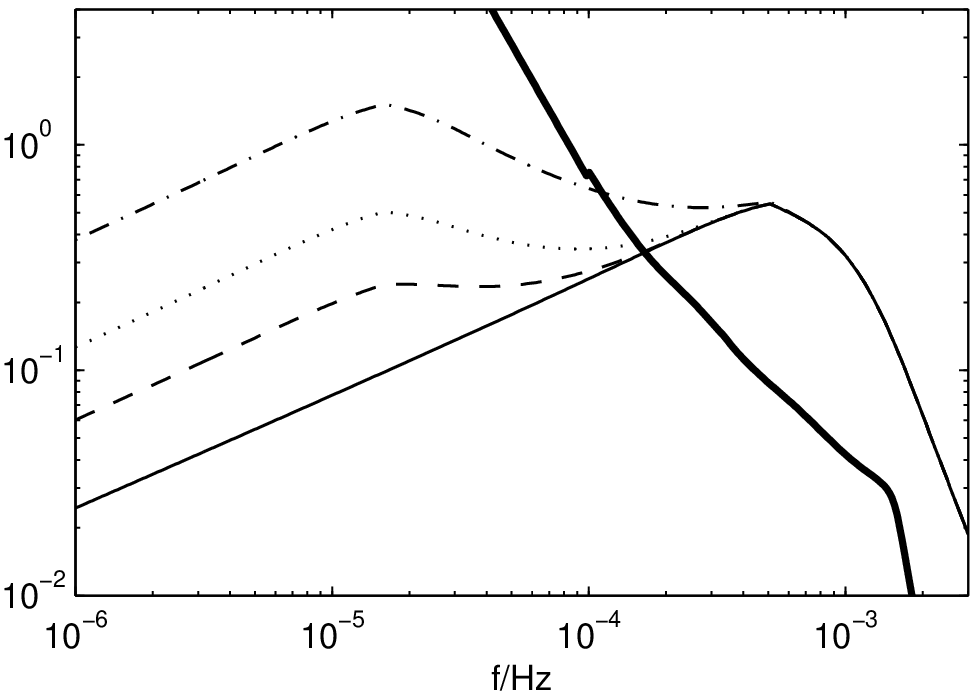}
\caption{The total spectrum of gravitational radiation  for
$g_\ast=100$, $T_\ast=100\,{\rm GeV}$, $\alpha_\ast=0.5$,
$\beta=100H_\ast$, and four different values of $\zeta_\ast$:
$\zeta_\ast=0$, corresponding to hydrodynamic turbulence without
the inverse cascade effect (solid line), $\zeta_*=0.02$ (dash
line), $\zeta_*=0.5$ (dotted line), and $\zeta_*=0.15$ (dot-dash
line), with the LISA sensitivity curve.} \label{fig:spectrum-zeta}
\end{figure}

Figure 1 displays the three components contributing to the total
gravitational wave spectrum, for a fiducial model with $g_*=100$,
$\alpha=0.5$, $\beta/H_*=100$, and $T_*=100$ GeV. Bubble
collisions are shown as the dashed line, hydrodynamic turbulence
with no magnetic helicity as the solid line, and MHD turbulence
with an initial helicity parameter $\zeta_*=0.15$. The dark curve
is the LISA design sensitivity curve for stochastic backgrounds
with a detection signal-to-noise ratio of 5 over a 1-year
integration \cite{curve}, including a rough estimate of the
confusion limit from white-dwarf binaries \cite{whitedwarfs}. It
is clear that bubble collisions alone do not produce a significant
signal (marginally detectable) in LISA for this range of phase
transitions parameters, agreeing with previous results
\cite{detection,nicolis,gs06,cds07}. The turbulence peak is
observable, as well as the separate peak from the MHD inverse
cascade.

\begin{figure}
\includegraphics[width=6.4cm]{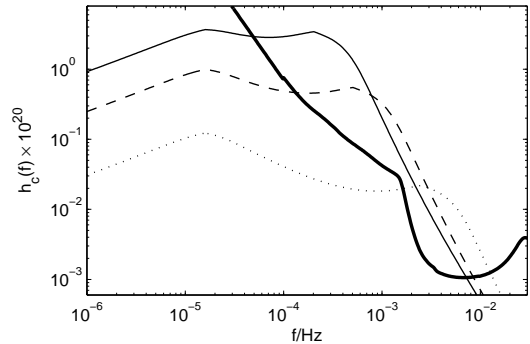}
\caption{The total spectrum of gravitational radiation  for
$g_\ast=100$, $T_\ast=100\,{\rm GeV}$, $\zeta_\ast=0.1$,
$\alpha=0.5$, and three different values of $\beta$:  $\beta=40
H_\ast$ (solid line), $\beta=100 H_\ast$ (dash line) and
$\beta=500 H_\ast$ (dotted line),  with the LISA sensitivity curve.} \label{fig:spectrum-beta}
\end{figure}

The next set of figures show how the total gravitational wave
spectrum varies as each of the phase transition parameters is
changed, holding the others fixed.
Figure~\ref{fig:spectrum-alpha} shows the variation with $\alpha$,
Fig.~\ref{fig:spectrum-T} the variation with $T_*$,
Fig.~\ref{fig:spectrum-zeta} the variation with $\zeta_*$, and
Fig.~\ref{fig:spectrum-beta} the variation with $\beta/H_\star$.
In each case, the spectrum for the fiducial model in Fig.~1 is
plotted, along with the spectra for two or three other values of
the parameter in question. We can see qualitative outlines of
detectability from these figures. For example,
Fig.~\ref{fig:spectrum-T} shows that as the temperature of the
phase transition increases, detection relies increasingly on the
MHD signal.

While these plots give a good qualitative sense of the spectrum
dependence on the parameters,  we would like to know more
precisely which regions of the four-dimensional parameter space of
phase transitions result in a gravitational wave spectrum
detectable with LISA. The following section gives one useful
representation of the detectability regions.

\section{Detectable Regions in Parameter Space}

\begin{figure}
\includegraphics[width=8.6cm]{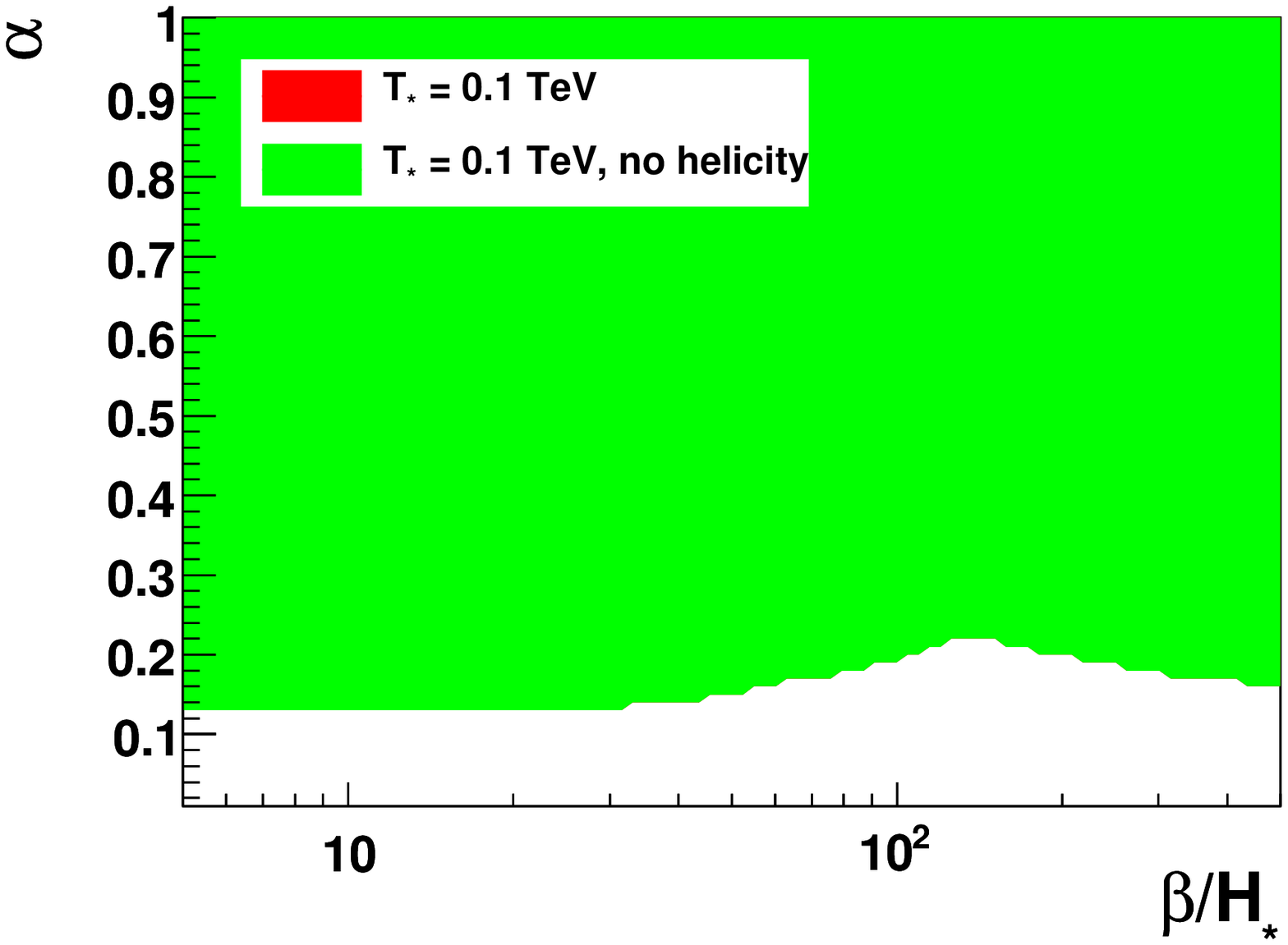}
\caption{The $\alpha$ vs.\ $\beta/H_\star$ LISA sensitivity region for
a phase transition at a temperature $T_*=0.1$ TeV. The regions for
$\zeta_*=0$ and $\zeta_*=0.15$ coincide at this temperature.
A point in parameter space is considered
detectable if at any frequency its value of $h_c(f)$ is detectable
at a signal-noise ratio of 5 in a one-year integration, including
the confusion noise from white dwarfs from \cite{curve}.}
\label{fig:sens_0.1TeV}
\end{figure}

\begin{figure}
\includegraphics[width=8.6cm]{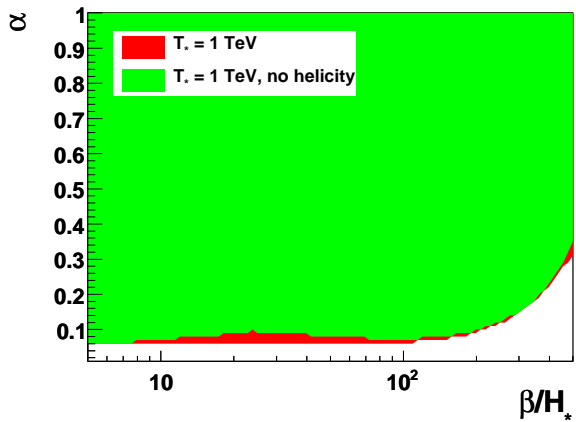}
\caption{Same as Fig.~\ref{fig:sens_0.1TeV}, except for
$T_*=1$ TeV. For $\zeta_*=0.15$, a slight additional region
along the low-$\alpha$ edge of the $\zeta_*=0$ region is
detectable.}
\label{fig:sens_1TeV}
\end{figure}

\begin{figure}
\includegraphics[width=8.6cm]{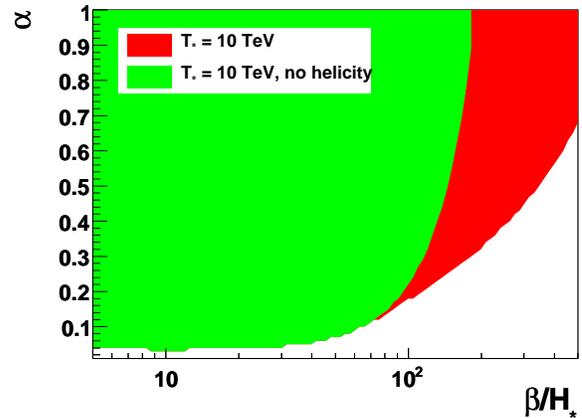}
\caption{Same as Fig.~\ref{fig:sens_0.1TeV}, except for
$T_*=10$ TeV. For $\zeta_*=0.15$, a substantial extra
area on the right side of the $\zeta_*=0$ region is detectable.}
\label{fig:sens_10TeV}
\end{figure}

\begin{figure}
\includegraphics[width=8.6cm]{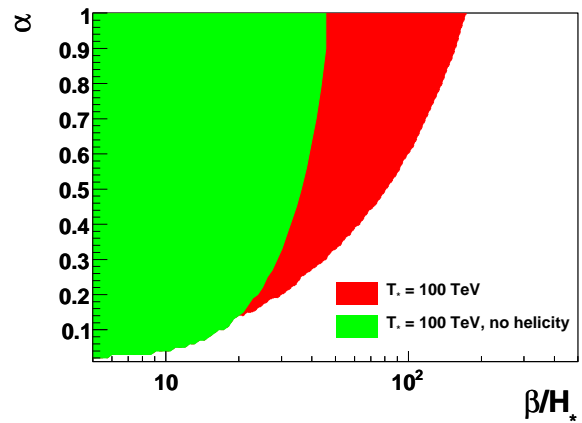}
\caption{Same as Fig.~\ref{fig:sens_10TeV}, except for $T_*=100$ TeV.}
 \label{fig:sens_100TeV}
\end{figure}

The remaining figures show various detectability regions. For the
first set of four plots, Figs.~\ref{fig:sens_0.1TeV} to
\ref{fig:sens_100TeV}, the detectable region in the
$\beta/H_\star$--$\alpha$ plane is displayed; each figure is for a
different value of the phase transition temperature $T_*$, ranging
from 0.1 TeV to 100 TeV. The larger displayed region is including
the MHD inverse cascade from an initial helicity of $\zeta=0.15$,
while the smaller region shows the case of no inverse cascade, or
equivalently the limit $\zeta=0$. The other parameter, $g_*$, is
held fixed and only has a weak effect on the results. We designate
a power spectrum as detectable if its amplitude comes above the
LISA 1-year sensitivity curve at signal-to-noise ratio of 5 at
any frequency; more careful detection conditions can also be used
but will give the same general results.

\begin{figure}
\includegraphics[width=8.6cm]{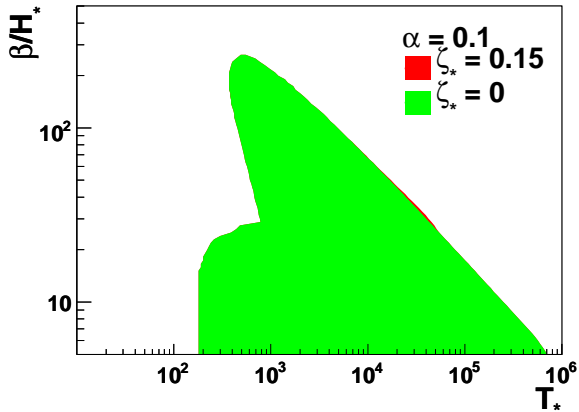}
\caption{The $\beta/H_*$ vs.\ $T_*$ LISA sensitivity region for
a phase transition with vacuum energy $\alpha=0.1$. The regions for
$\zeta_*=0$ and $\zeta_*=0.15$ coincide at this temperature.
A point in parameter space is considered
detectable if at any frequency its value of $h_c(f)$ is detectable
at a signal-noise ratio of 5 in a one-year integration, including
the confusion noise from white dwarfs from \cite{curve}.}
\label{fig:sens_betaVsT_alpha0.1}
\end{figure}

\begin{figure}
\includegraphics[width=8.6cm]{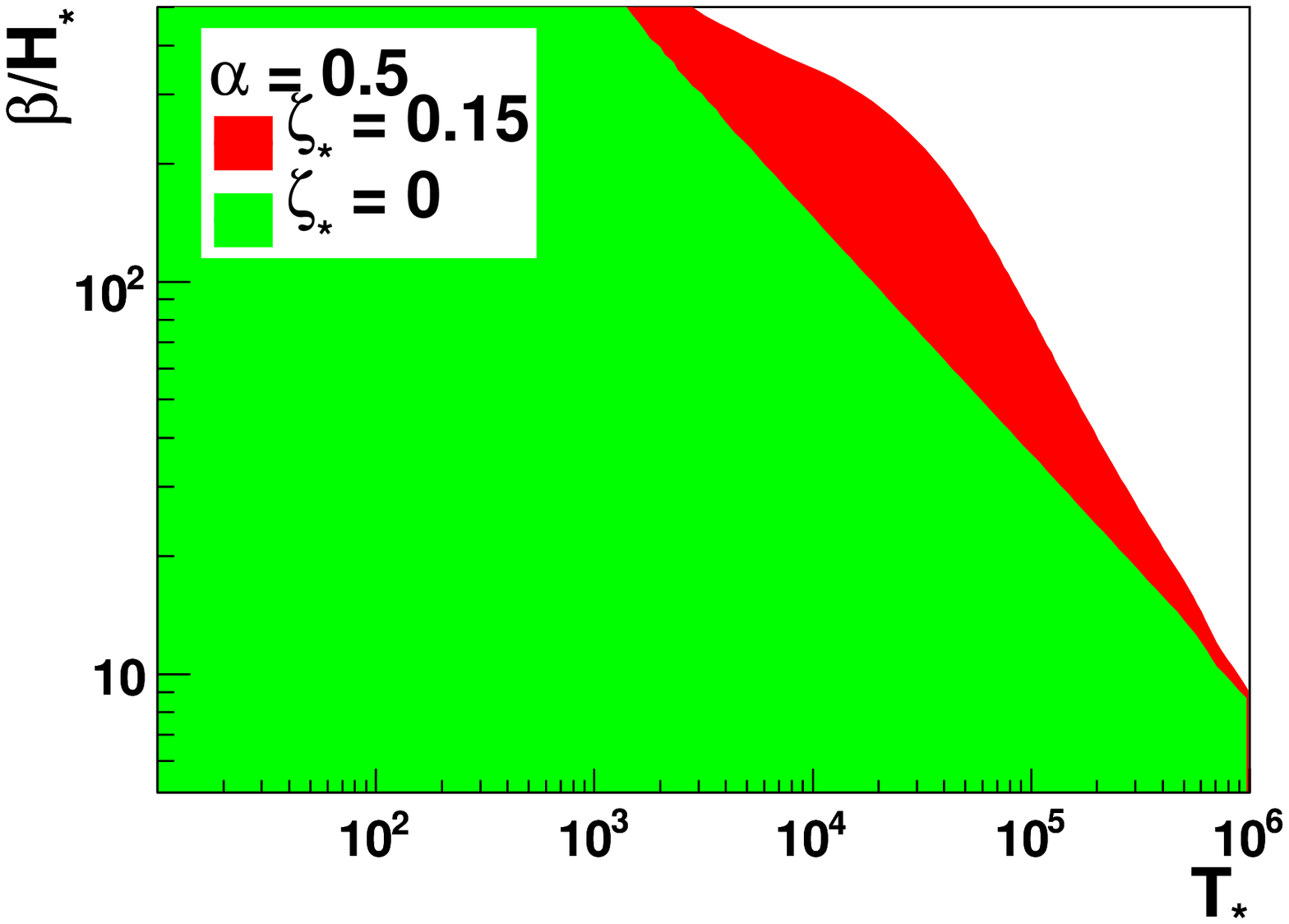}
\caption{Same as Fig.~\ref{fig:sens_betaVsT_alpha0.1}, except for $\alpha=0.5$.}
\label{fig:sens_betaVsT_alpha0.5}
\end{figure}

Large regions of parameter space are detectable. The natural
expectation for $\beta$ is around $100 H_*$ \cite{tww92}, with
some range around this value for plausible models. For
$\beta/H_\star$ between 10 and 100, the detection threshold in
$\alpha$ tends to be roughly constant, with lower $\alpha$ being
detectable as $T_*$ increases. If $T_* > 1$ TeV, then phase
transitions with $\alpha \geq 0.06$ are detectable by LISA.

For temperatures $T_* \leq 1$ TeV, the contribution from the MHD inverse cascade
is not important for LISA detectability, due to the experiment's low-frequency
cutoff. As $T_*$ moves to higher temperatures, the typical frequency, set
by $H_*$, becomes larger, and the low-frequency peak and tail from the MHD
inverse cascade starts to come into the detection window. For $T_*\geq 10$ TeV, the
detectability range in $\beta$ is substantially expanded due to the magnetic source,
though mostly at comparatively large values of $\alpha$. The spectrum contribution from
hydrodynamic turbulence, due to its somewhat higher frequency and amplitude than
that from bubble collisions and much higher frequency than the MHD contribution,
dominates most of the detectability for the LISA frequency band.

Figures \ref{fig:sens_betaVsT_alpha0.1} and \ref{fig:sens_betaVsT_alpha0.5} show the detectability regions in the $\beta$--$T_*$ plane,
for $\alpha=0.1$ and $\alpha=0.5$, respectively. For $\alpha=0.1$, a relatively
large region is detectable, while for $\alpha=0.5$, a phase transition at any temperature
up to 10 TeV is detectable for any reasonable value of $\beta$.

These results will be useful for computing future constraints on various high-energy
physics theories which give first-order electroweak phase transitions. The fundamental
parameters of any theory can be used to construct order parameters and vacuum energies
for any phase transition at relevant energy scales (see, e.g., Refs.~\cite{strong} for many example).
These in turn can be used to compute the phenomenological phase transition parameters
$\alpha$, $\beta$, and $T_*$ used here, as done in Appendix A of Ref.~\cite{kos1}.
The parameter $\zeta_*$ likely will require
additional astrophysical modelling in situations where there is no well-defined mechanism
for generating magnetic fields, but most of the parameter-space detectability is independent
of this parameter.

We have made several basic physical assumptions: we have assumed that
the walls of the bubbles formed in the first order phase transition expand as detonations
\cite{kf92}; that the bubbles remain spherical as they expand; that the resulting bubble
collisions source turbulence on the scale of the largest bubbles for a time comparable
to the duration of the phase transition; that the turbulence can be modeled as fully-developed
Kolmogoroff turbulence; and that initial seed magnetic fields with non-zero helicity will
be amplified in an inverse cascade according to a specific MHD turbulence model \cite{BM99,CHB05}.
While these are all reasonable assumptions and are likely close to the realistic physical
situation, they are not guaranteed to be correct. In particular, the assumption that bubbles in
a relativistic plasma always expand as detonations deserves further study. The transition
from coherent bubble wall expansion to stochastic turbulent behavior needs to be
simulated numerically; current computing power is sufficient to model this
process with enough resolution to verify the simple assumptions in this work.
The standard Kolmogoroff turbulence results apply to non-relativistic fluids, and
our direct extension to relativistic plasmas, while likely qualitatively correct, is not based on
either direct observation or detailed calculation. Finally, mechanisms for generating seed
magnetic fields prior to or during the phase transition are not clearly understood.
The enticing prospect of a direct probe of electroweak physics via the stochastic background
of gravitational radiation is a strong motivation for further investigation of these interesting
questions concerning the fundamental physics of the early universe.

 \acknowledgments We greatly appreciate useful
comments and discussions from L. Campanelli, G. Gabadadze, and
B.~Ratra. We thank the anonymous referee for spotting a numerical
error in an earlier draft of the paper. G.G. and T.K. acknowledge
partial support from grant 061000017-9258 from The International
Association for the Promotion of Cooperation with Scientists from
the Newly Independent States of the Former Soviet Union (INTAS),
from grant ST06/4-096 of the Georgian National Science Foundation
(GNSF), and from the Associate Member Program of the
International Center for Theoretical Physics (ICTP). T.K. and
Y.M. have received partial support from grant E-FG02-99ER41093
from the Department of Energy. A.K. has been partly supported
from National Science Foundation grant AST-0546035. G.G.
acknowledges partial support by the joint Award 06-11 from The
Georgian Research and Development Foundation (GRDF) and the U.S.\
Civilian Research and Development Foundation (CRDF), and grant
ST07/4-193 from the Georgian National Science Foundation (GNSF).

%\end{document}
%\newpage

%%%%%%%%%%%%%%%%%%


\begin{thebibliography}{}

\bibitem{inflation} L.P.~Grishchuk, Sov.\ Phys.\ JETP {\bf 40}, 409 (1975)
  [Zh.\ Eksp.\ Teor.\ Fiz.\  {\bf 67}, 825 (1974)];
A.~Starobinsky, Sov.\ Phys.\ JETP Lett.\ {\bf 30}, 682 (1979)
[Pisma Zh.\ Eksp.\ Teor.\ Fiz.\ {\bf 30}, 719 (1979)]; V.A.~Rubakov,
M.V.~Sazhin, and A.V.~Veryaskin,
 Phys.\ Lett.\ B {\bf 115}, 189 (1982); B.~Allen, Phys.\ Rev.\ D
{\bf 37}, 2078 (1988); B.~Ratra, Phys.\ Rev.\ D {\bf 45} 1913
(1992); M.~Giovannini, Phys.\ Rev.\ D {\bf 60}, 123511 (1999).

\bibitem{end-inflation} S.Y.~Khlebnikov and I.I.~Tkachev, Phys.\ Rev.\ D {\bf 56}, 653 (1997);
R.~Easther and E.A.~Lim, J.\ Cosm.\ Astropart.\ Phys.\  {\bf 0604}, 010 (2006);
J.~Garcia-Bellido and D.G.~Figueroa, Phys.\ Rev.\ Lett.\ {\bf
98}, 061302 (2007). R. Easther, J. T. Giblin, and E. A. Lim,
arXiv:0712.2991 [astro-ph]; J.~Garcia-Bellido, D.G.~Figueroa and
A.~Sastre, Phys.\ Rev.\  D {\bf 77}, 043517 (2008); A.~Megevand,
arXiv:0804.0391 [astro-ph]. .

\bibitem{bubble} E.~Witten, Phys.\ Rev.\ D {\bf 30}, 272 (1984); C.J.~Hogan, Mon.\ Not.\ R.\
Astron.\ Soc.\ {\bf 218}, 629 (1986);  M.S.~Turner and
F.~Wilczek, Phys.\ Rev.\ Lett.\ {\bf 65}, 3080 (1990).

\bibitem{kos1}
M.~Kamionkowski, A.~Kosowsky, and M.\ S.\ Turner, Phys.\ Rev.\ D
{\bf 49}, 2837 (1994).

\bibitem{ktw92a} A.~Kosowsky, M.S.~Turner, and R.~Watkins,
    Phys.\ Rev.\ D {\bf 45}, 4514 (1992).

\bibitem{kt93} A.~Kosowsky and M.S.~Turner, Phys.\ Rev.\ D
    {\bf 47}, 4372 (1993); A~Kosowsky, M.S.~Turner, and R.~Watkins,
    Phys.\ Rev.\ Lett.\ {\bf 69}, 2026 (1992).

\bibitem{kos2} R.~Apreda, et al., Class.\ Quant.\ Grav. {\bf 18},  L155 (2001);
J.F.~Dufaux, et al., Phys.\ Rev.\ D {\bf 76}, 123517 (2007);
S.J.~Huber and T.~Konstandin, arXiv:0709.2091 [hep-ph];
S.~J.~Huber and T.~Konstandin,
  %``Gravitational Wave Production by Collisions: More Bubbles,''
  arXiv:0806.1828 [hep-ph].

\bibitem{strings}  T.~Vachaspati and A.~Vilenkin,
  Phys.\ Rev.\  D {\bf 31}, 3052 (1985);  M.P.~Infante and N.~S\'anchez, Phys.\ Rev.\ D
{\bf 61}, 083515 (2000); S. G. Rubin, A.S.Sakharov, M.Yu.
Khlopov, J.\ Exp.\ Theor.\ Phys.\  {\bf 91}, 921 (2001)
  [J.\ Exp.\ Theor.\ Phys.\  {\bf 92}, 921 (2001)]
  [arXiv:hep-ph/0106187].
  .

\bibitem{magnet} D.V.~Deriagin, et al., Mon.\ Not.\ R.\ Astron.\ Soc.\ {\bf 229}, 357
(1987); R.~Durrer, P.~Ferreira, and T.~Kahniashvili, Phys.\ Rev.\ D
 {\bf 61}, 043001 (2000);
 A.~Mack, T.~Kahniashvili, and A.~Kosowsky, Phys.\ Rev.\ D {\bf 65}, 123004 (2002);
A.~Lewis, Phys. \ Rev. \ D. {\bf 70}, 043011 (2004).

\bibitem{cdk04} C.~Caprini, R.~Durrer, and T.~Kahniashvili,
Phys.\ Rev.\ D {\bf 69}, 063006 (2004).

\bibitem{cd06} C.~Caprini and R.~Durrer, Phys.\ Rev.\ D {\bf 74},
063521 (2006).

\bibitem{kmk02} A.~Kosowsky, A.~Mack, and T.~Kahniashvili,
Phys.\ Rev.\ D {\bf 66}, 024030 (2002).

\bibitem{dolgov} A.D.~ Dolgov, D.~Grasso, and A.~Nicolis,
 Phys.\ Rev.\ D {\bf 66}, 103505 (2002).

\bibitem{kgr05} T.~Kahniashvili, G.~Gogoberidze, and B.~Ratra,
Phys.\ Rev.\ Lett.\ {\bf 95}, 151301  (2005).

\bibitem{gkk07} G.~Gogoberidze, T.~Kahniashvili, and A.~Kosowsky,
Phys.\ Rev.\ D {\bf 76}, 083002 (2007).

\bibitem{lisa} http://lisa.nasa.gov/

\bibitem{detection}  R.~Apreda et al., Nucl.\ Phys.\ B {\bf 631}, 342 (2002).

\bibitem{nicolis} A.~Nicolis, Class. \ Quant.\  Grav.\ {\bf 21}, L27 (2004).

\bibitem{gs06} C.~Grojean and G.~Servant, Phys. \ Rev. \ D {\bf 75} 043507 (2007).

\bibitem{grojean} C.~Delaunay, C.~Grojean and J.~D.~Wells,
  %``Dynamics of Non-renormalizable Electroweak Symmetry Breaking,''
  J.\ High Energy Phys.\ {\bf 0804}, 029 (2008);

\bibitem{stdmodelweak}  K.~Kajantie, M.~Laine, K.~Rummukainen, and M.~Shaposhnikov,
  Phys.\ Rev.\ Lett.\ {\bf 77}, 2887 (1996);
  M.~Gurtler, E.M.~Ilgenfritz, and A.~Schiller, Phys.\ Rev.\ D {\bf 56}, 3888 (1997).

 \bibitem{strong} G.W.~Anderson and L.J.~Hall, Phys.\ Rev.\ D {\bf 45}, 2685 (1992);
  M.~Laine and K.~Rummukainen, Phys.\ Rev.\ Lett.\ {\bf 80}, 5259 (1998);
  S.J.~Huber and M.G.~Schmidt, Eur.\ Phys.\ J.\ {\bf C10}, 473 (1999);
  S.W.~Ham and S.K.~Oh, Phys.\ Rev.\ D {\bf 70}, 093007 (2004);
  C.~Grojean, G.~Servant, and J.~Wells, Phys.\ Rev.\ D {\bf 71}, 036001 (2005);
S.~Huber et al., Nuc.\ Phys.\ {\bf B757}, 172 (2006);
L.~Randall and G.~Servant,
%``Gravitational Waves from Warped Spacetime,''
J.\ High Energy Phys.\ {\bf 0705}, 054 (2007).

\bibitem{helicity} J.~Cornwall, Phys.\ Rev.\ D {\bf 56}, 6146 (1997);
M.~Giovannini and M.~E.~Shaposhnikov, Phys.\ Rev.\  D {\bf 57}, 2186 (1998);
G.B.~Field and S.M.~Carroll, Phys.\ Rev.\  D {\bf 62}, 103008 (2000);
T.~Vachaspati, Phys.\ Rev.\ Lett.\ {\bf 87}, 251302 (2001);
G.~Sigl, Phys.\ Rev.\  D {\bf 66}, 123002 (2002);
K.~Subramanian and A.~Brandenburg,  Phys.\ Rev.\ Lett.\ {\bf 93}, 205001 (2004);
L.~Campanelli and M.~Giannotti, Phys.\ Rev.\  D {\bf 72}, 123001 (2005);
V.B.~Semikoz and D.D.~Sokoloff, Astron.\ Astrophys.\ {\bf 413}, L53 (2005);
A.~Diaz-Gil, J.~Garcia-Bellido, M.~Garcia-Perez and
A.~Gonzalez-Arroyo,
  %``Magnetic field production during preheating at the electroweak scale,''
  arXiv:0712.4263 [hep-ph]; L.~Campanelli,
  %``Helical Magnetic Fields from Inflation,''
  arXiv:0805.0575 [astro-ph].
\bibitem{ste08} T.~Stevens et al., Phys.\ Rev.\ D {\bf 77}, 3501 (2008).

\bibitem{kgr08} T.~Kahniashvili, G.~Gogoberidze, and B.~Ratra,
arXiv:0802.35245 [astro-ph], Phys. Rev. Lett., in press

\bibitem{seto} N.~Seto, Phys.\ Rev.\ Lett.\ {\bf 97}, 151101
(2006).

\bibitem{m00} M.~Maggiore, Phys.\ Rep.\ {\bf 331}, 28 (2000).

\bibitem{W} S.~Weinberg, {\sl Gravitation and Cosmology}
(Wiley \& Sons, New York, 1972).

\bibitem{gs07} C.~Delaunay, C.~Grojean and G.~Servant,
 %``The Higgs in the sky: Production of gravitational waves during a
 %first-order phase transition,''
AIP Conf.\ Proc.\  {\bf 903}, 24 (2007).

\bibitem{cds07} C.~Caprini, R.~Durrer, and G.~Servant, arXiv:0711.2593
[astro-ph].

\bibitem{tww92} M.S.~Turner, E.J.~Weinberg, and L.M.~Widrow,
      Phys.\ Rev.\ D {\bf 46}, 2384 (1992).

\bibitem{col77} S.~Coleman, Phys.\ Rev.\ D {\bf 15}, 2929 (1977);
C.G.~Callan and S.~Coleman, Phys.\ Rev.\ D {\bf 16}, 1762 (1977).

%\bibitem{ghk97} M.~Gleiser, A.F.~Heckler, and E.W.~Kolb,
%Phys.\ Lett.\ B {\bf 405}, 121 (1997).

\bibitem{ste82} P.J.~Steinhardt, Phys.\ Rev.\ D {\bf 25}, 2082 (1982).

\bibitem{kf92} M.~Kamionkowski and K.~Freese, Phys.\ Rev.\ Lett.\ {\bf 69}, 2743 (1992).

\bibitem{G} M.E.~Goldstein, {\sl Aeroacoustics} (McGraw-Hill, New York, 1976).

\bibitem{B03} D.~Biskamp, {\sl Magnetohydrodynamic Turbulence}
(Cambridge University, Cambridge, 2003); M.K.~Verma, Phys.\ Rep.\
{\bf 401}, 229 (2004).


\bibitem{my75} A.S.~Monin and A.M.~Yaglom, {\sl Statistical Fluid Mechanics}
(MIT Press, Cambridge, MA, 1975).

\bibitem{K64} R.H.~Kraichnan, Phys.\ Fluids {\bf 7}, 1163 (1964).

\bibitem{BM99} D.~Biskamp and W.C.~Mueller, Phys.\ Rev.\ Lett.\ {\bf 83}, 2195 (1999);
Phys.\ Plasma {\bf 7}, 4889 (2000).

\bibitem{CHB05}
M.~Christensson, M.~Hindmarsh, and A.~Brandenburg, Phys.\ Rev.\ E
{\bf 64}, 056405 (2001); Astron.\ Nachrichten {\bf 326}, 393
(2005).

\bibitem{jedamzik} R.~Banerjee and K.~Jedamzik, Phys.\ Rev.\ D {\bf 70}, 123003
(2004); L. Campanelli, Phys. Rev. Lett. {\bf 98}, 251302 (2007).

\bibitem{klgr08a} T.~Kahniashvili, L. Campanelli, G.~Gogoberidze, and B.~Ratra, in
preparation

\bibitem{curve}
S.L.~Larsen, W.A.~Hiscock, and R.W.~Hellings, Phys.\ Rev.\ D {\bf 62}, 062001 (2000);
N.~Cornish, Phys.\ Rev.\ D {\bf 65}, 022004 (2001);
For a LISA sensitivity curve calculator used here, see
http://www.srl.caltech.edu/$\sim$shane/sensitivity/

\bibitem{whitedwarfs}
P.L.~Bender and D.~Hils, Cl.\ Quant.\ Grav.\ {\bf 14}, 1439 (1997).
















%\bibitem{gw-pol} A.~Lue, L.M.~Wang, and M.~Kamionkowski,
%Phys.\ Rev.\ Lett.\ {\bf 83}, 1506 (1999); D.~Lyth, C.~Quimbay,
%and Y.~Rodriguez, J.\ High Energy Phys.\  {\bf 0503}, 016 (2005); M.~Satoh, S.~Kanno,
%and J.~Soda, Phys.\ Rev.\ D {\bf 77}, 023526 (2008).

%
%\bibitem{a} S.H.S.~Alexander, M.E.~Peskin, and M.M.~Sheikh-Jabbari,
%Phys.\ Rev.\ Lett.\ {\bf 96}, 081301 (2006); S.~Saito, K.~Ichiki,
%and A.~Taruya, J.\ Cosm.\ Astropart.\ Phys.\ {\bf 0709}, 002 (2007).

%

%

%%\bibitem{L52} M. J. Lighthill, Proc. R. Soc. London Ser. A {\bf
%%211}, 564 (1952), {\bf 222}, 1 (1954).

%%\bibitem{P52} I. Proudman, Proc. R. Soc. London Ser. A {\bf 214,}
%%119 (1952).

%
%%\bibitem{s83} J. V. Shebalin, W. H. Matthaeus, and D. Montgomery, J. Plasma
%%Phys. {\bf 29,} 525 (1983); P. Goldreich and S. Sridhar,
%%Astrophys. J. {\bf 438,} 763 (1995).

%\bibitem{mtw73} C.~Misner, K.S.~Thorne, and J.A.~Wheeler,
%{\sl Gravitation} (W.H.~Freeman, San Francisco, 1973), Sec.\ VIII.

%\bibitem{kol41} A. N.~Kolmogorov, Dokl.\ Akad.\ Nauk.\ SSSR
%{\bf 30}, 299 (1941).


%\bibitem{meszaros} S.~Kobayashi and P.~M{\'e}sz{\'a}ros,
% Astrophys.~J.~Lett.~{\bf 585}, L89 (2003).

\end{thebibliography}
\end{document}